\documentclass[sigconf]{acmart}

\renewcommand\footnotetextcopyrightpermission[1]{} 
\usepackage{booktabs} 
\usepackage{hyperref}
\usepackage{algorithm}
\usepackage{algpseudocode}
\usepackage{amsmath}
\usepackage{amssymb}
\usepackage[utf8]{inputenc} 
\usepackage[T1]{fontenc}    
\usepackage{hyperref}       
\usepackage{url}            
\usepackage{booktabs}       
\usepackage{amsfonts,amsmath,amssymb}       
\usepackage{nicefrac}       
\usepackage{microtype}      
\usepackage{graphicx}
\usepackage{multirow}
\usepackage{subcaption}
\usepackage{amsfonts}

\begin{document}
\title{Learning Image Information for eCommerce Queries}

\author{Utkarsh Porwal}
\affiliation{%
  \institution{eBay Inc}
  \city{San Jose}
  \state{California}
}
\email{uporwal@ebay.com}
%

\begin{abstract}
Computing similarity between a query and a document is fundamental in any information retrieval system. In search engines, computing query-document similarity is an essential step in both retrieval and ranking stages. In eBay search, document is an item and the query-item similarity can be computed by comparing different facets of the query-item pair. Query text can be compared with the text of the item title. Likewise, a category constraint applied on the query can be compared with the listing category of the item. However, images are one signal that are usually present in the items but are not present in the query. Images are one of the most intuitive signals used by users to determine the relevance of the item given a query. Including this signal in estimating similarity between the query-item pair is likely to improve the relevance of the search engine. We propose a novel way of deriving image information for queries. We attempt to learn image information for queries from item images instead of generating explicit image features or an image for queries. We use canonical correlation analysis (CCA) to learn a new subspace where projecting the original data will give us a new query and item representation. We hypothesize that this new query representation will also have image information about the query. We estimate the query-item similarity using a vector space model and report the performance of the proposed method on eBay's search data. We show 11.89\% relevance improvement over the baseline using area under the receiver operating characteristic  curve (AUROC) as the evaluation metric. We also show 3.1\% relevance improvement over the baseline with area under the precision recall curve (AUPRC) . 
\end{abstract}

\keywords{CCA; Query-Item Similarity; Vector Space Model; Canonical Correlation Analysis}

\maketitle

\section{Introduction}

In e-commerce search, estimating query-document similarity is essential for retrieval and ranking. Using vector space model both queries and the documents can be represented as feature vectors. Similarity between their corresponding feature vectors would act as a proxy for the query-document similarity. Similarity between the query-document pair can either be fed as one of the many features in training a machine learned ranker. Or it can be used as a stand alone measure for relevance. Note in this work we assume queries as text queries issued with or without any category constraint to limit the search only to that category. Likewise, documents are items represented by item title, item image and the listing category\footnote{e-commerce sites commonly group listed items into a browsable set of categories.} of the item. We do not consider any other source of information such as price or description of the item. To estimate query-item similarity different features are extracted from different facets of the query and the item. For instance, text of the query can be compared with the text of the item title. Likewise, category constraint on the query can be compared with the listing category of the item. However, one key source of information missing in the queries that every item has is the image. There is no image associated with the query and it is also not easy to infer one because of challenges outlined in the Section ~\ref{sec:related}.

In this paper, we propose a novel method to derive image information for queries. We use canonical correlation analysis (CCA) to learn a latent subspace where both query projection and item projection will be most predictive of each other. We hypothesize that after projecting on the learned subspace, query vector will encapsulate information about images as well. Note that we are not learning explicit image features for queries. We are learning a multi-view representation of the queries and the items where they will have all the information available about one another that would help capture the similarity between them. Using vector space model, we will estimate this similarity using the cosine between the query-item vectors. We first propose a baseline cosine similarity between the query-item feature vectors that does not have any image information in them. We compare this baseline with our proposed method where we include the image information about both query and the item using the item images and show significant improvement in improving the relevance of the vector space model.

The rest of the paper is organized as follows. Section~\ref{sec:related} describes the related research in the area of canonical correlation analysis and image generation conditioned on text. Section~\ref{sec:feat} covers the details of features used in this work. Section~\ref{sec:cca} explains our proposed method where we discuss how to use CCA. Data sets descriptions and results are outlined in Section~\ref{sec:exp}. In Section~\ref{sec:conclusion} we highlight the main contribution of this work.

\section{Related Work}
\label{sec:related}
Canonical correlation analysis (CCA) \cite{cca}\cite{borga} is a powerful statistical technique of correlating linear
relationships between two multidimensional variables. Simply put if there are two data views of the same object then CCA will learn a latent space such that each view's
representation is the most predictive of the other and vice versa. It is commonly used for multi view representation learning \cite{nips2017}\cite{ge2016}. Different variants have been proposed to extend the capabilities of CCA. Kernel canonical correlation analysis (KCCA) is a nonparametric method that extends CCA to learn non linear correlated transformation\cite{kcca}\cite{kcca2}. Andrew et al. \cite{dcca} proposed deep canonical correlation (DCCA) analysis which addressed the scalability concerns about KCCA. It was a parametric method and authors showed the efficacy of their algorithm on handwritten digit recognition dataset and speech dataset. Closest to our application is the work by Yan et al. \cite{yan} where they used deep canonical correlation (DCCA) for matching images with captions. However, queries are significantly shorter than captions and descriptions, learning latent representation for query and item images is extremely challenging.

Recently there has been work in generating images given text. It falls under the multimodal learning framework where one modality has to be learned conditioned on another modality. Image generation given text is particularly difficult because there could be many possible pixel configurations that satisfy the given description (text). Scott et al. \cite{reed} proposed a text to image synthesis approach using a deep convolutional generative
adversarial network (DC-GAN) conditioned on text features. Later, Zhang et al. \cite{stackgan} proposed a StackGAN approach to generate high quality realistic images from text description using stacked generative adversarial networks. However, both of these approaches used long textual description to generate images. These methods suffer from the same problems as CCA methods is that they cannot be easily extended to search applications where queries are usually short. Queries are not descriptive in a traditional natural language sense. Moreover, e-commerce queries often consist of a bag of independent words and phrases identifying desired attributes. For example, brand and product names, color, sizes and units of measures. This makes image generation a significantly challenging problem for e-commerce queries. In this work we attempt to learn image information from item images instead of explicitly learning image features or generating an image for queries. 

\section{Features Used}
\label{sec:feat}
We use vector space model to represent queries and the items as vectors. We derive a hashed \textit{tf-idf} vector of both query text and the item title. We also extract the image features from the item image. We would concatenate the hashed \textit{tf-idf} vector of item with its image feature vector. Since the query-item similarity is defined as the cosine similarity between the query vector and the item vector, it is imperative for both the vectors to be equidimensional. However, in the absence of the image information for queries, query-item vectors are not  equidimensional. We will use CCA to get equidimensional feature vectors of both query and the item by projecting them to a learned subspace where cosine similarity between the projected vectors can be computed. We present the details of feature extraction in the following subsections.

\subsection{Hashed \textit{TF-IDF}}
There are different ways to generate \textit{tf-idf} vectors for the query and the item title. One option is to consider all the items in the index and generate  \textit{tf-idf} vectors of the size of the vocabulary of the index. With this approach there is a risk of generating sparse vectors and diluting the information captured in the feature vector as e-commerce sites often have a wide variety of items in their index. Instead, we used listing category of the item to generate the \textit{tf-idf} vector for the item. For every query-item pair, we considered all the item titles present in the listing category of the corresponding item to generate the \textit{tf-idf} vectors for both query and the item. We made an assumption that both query and item title come from the same distribution. In reality these distributions can be different, but in this work we are ignoring this effect. This is to simplify the \textit{tf-idf} vector generation process and so that we can objectively evaluate the effect of adding image signal by keeping that as the only major differentiating factor. The impact of distribution differences can be explored in follow-on studies.

Since we formulated the query-item similarity as vector similarity, \textit{tf-idf} vectors for the query-item pair should be equidimensional. Our current approach would result in different sizes of query-item vectors depending on the vocabulary size of the corresponding listing category of the item title. Note that  vocabulary size would explode if bi-grams, tri-grams and/or skip grams are considered as part of the vocabulary. This creates another issue of keeping the mapping of words in the vocabulary to their corresponding feature index in memory. This memory constraint gets accentuated with the fact that we have to maintain such mapping for every listing category. eBay has over $15,000$ listing categories. Therefore, we applied the hashing trick \cite{hashing} on the \textit{tf-idf} vectors of both query and the item title which addresses all such concerns. It \textit{hashes} the high dimensional \textit{tf-idf} vector $\mathbf{x}$ to a lower dimensional feature space of $d$ dimensions. However, this benefit of equidimensional vectors with low memory footprint comes at a cost. There is a possibility of collision when more than one word in the vocabulary maps to the same index. This probability depends on the dimensionality of the lower dimensional vector. Keeping this in mind, we optimized the number of collisions against the different vocabulary size of different categories and created all the \textit{tf-idf} vectors of $1000$ dimensions. Other drawback of hashing is the inability to get the original representation back as the \textit{idf} weights are not stored as that would mean a higher memory footprint. 


\subsection{Image Features}
We trained a Resnet-50 model \cite{resnet} for image to category prediction and used the features from the layer before the prediction layer. In eBay's category tree we have over 15,000 listing categories and every item is listed in at least one listing category. We collected around 5000 images per listing category for training the model and around 200 images per listing category for validation. This model is used for several applications that would require image to category prediction. We query the trained model with item image and get a $1024$ dimensional vector from the layer before the prediction layer.

\section{Proposed Approach}
\label{sec:cca}

For \textit{t} query-item pairs, we have item image features $\mathbf{U} \in \mathbb{R}^{t \times u}$ and \textit{tf-idf} features $\mathbf{V} \in \mathbb{R}^{t \times v}$ for the item titles representing items. An item can be represented as a random variable $\mathbf{I} \in \mathbb{R}^{t \times n}$ where
$\mathbf{I}^\intercal = \begin{bmatrix}
 \mathbf{U} \\ 
 \mathbf{V}
 \end{bmatrix}$ with $\textit{n} = \textit{u} + \textit{v}$.
  Likewise, \textit{tf-idf} features for query can be represented as random variable $\mathbf{Q} \in \mathbb{R}^{t \times m}$. We would like to learn a common subspace between $\mathbf{I}$ and $\mathbf{Q}$ so that we can calculate the similarity between the two in the learned subspace. Canonical Correlation Analysis (CCA) is a statistical technique that learns a linear relationship between two multidimensional variables such as $\mathbf{I}$ and $\mathbf{Q}$. It will learn a set of basis vectors for each $\mathbf{I}$ and $\mathbf{Q}$ such that the correlations between the projections of the variables onto these basis vectors are mutually maximized \cite{borga}. Given that random variable $\mathbf{Q}$ and $\mathbf{I}$ have zero mean, we can write the total covariance matrix as 

\begin{displaymath}
\mathbf{C} = \begin{bmatrix}
 \mathbf{C}_{\mathbf{Q}\mathbf{Q}} & \mathbf{C}_{\mathbf{Q}\mathbf{I}} \\ 
 \mathbf{C}_{\mathbf{I}\mathbf{Q}} & \mathbf{C}_{\mathbf{I}\mathbf{I}}
 \end{bmatrix}
= \mathbb{E} \begin{bmatrix}
 \binom{\mathbf{Q}}{\mathbf{I}} & \binom{\mathbf{Q}}{\mathbf{I}}^\intercal
 \end{bmatrix}
 \end{displaymath}

Here, $\mathbf{C}_{\mathbf{Q}\mathbf{Q}}$ and $\mathbf{C}_{\mathbf{I}\mathbf{I}}$ are the within set covariance matrices of query and image features respectively and $\mathbf{C}_{\mathbf{Q}\mathbf{I}}$, $\mathbf{C}_{\mathbf{I}\mathbf{Q}}$ are the between set covariance matrices where 

\begin{displaymath}\mathbf{C}_{\mathbf{Q}\mathbf{I}} = \mathbf{C}_{\mathbf{I}\mathbf{Q}}^\intercal \end{displaymath}

As $\mathbf{Q}$ is \textit{m} dimensional long and $\mathbf{I}$ is \textit{n} dimensional long with \textit{m} < \textit{n}, CCA would learn \textit{m} basis vectors for query and item variables respectively. Let us take a case where each set would only have one basis vector i.e we want to learn only one pair of basis vectors (canonical variate pair) with the largest canonical correlation. Given the linear combination of query as $Q=\mathbf{Q}^\intercal\hat{\mathbf{W}}_{Q}$ and item as $I=\mathbf{I}^\intercal\hat{\mathbf{W}}_{I}$, canonical correlation is defined as  
 
\begin{equation} \rho =  \frac{\mathbb{E}[QI]}{\sqrt{\mathbb{E}[Q^2]E[I^2]}} = \frac{\mathbb{E}[\hat{\mathbf{W}}_{Q}^{\intercal}\mathbf{Q}\mathbf{I}^\intercal\hat{\mathbf{W}}_{I}]}{\sqrt{\mathbb{E}[\hat{\mathbf{W}}_{Q}^{\intercal}\mathbf{Q}\mathbf{Q}^\intercal\hat{\mathbf{W}}_{Q}]\mathbb{E}[\hat{\mathbf{W}}_{I}^{\intercal}\mathbf{I}\mathbf{I}^\intercal\hat{\mathbf{W}}_{I}]}}  \end{equation}

\begin{equation} = \frac{\mathbf{W}_{Q}^\intercal\mathbf{C}_{QI}\mathbf{W}_{I}}{\sqrt{\mathbf{W}_{Q}^\intercal\mathbf{C}_{QQ}\mathbf{W}_{Q}\mathbf{W}_{I}^\intercal\mathbf{C}_{II}\mathbf{W}_{I}}} \end{equation}

For consecutive pair of basis vectors we would have further constraints. $\forall i \neq j $ 

 \begin{equation} \mathbb{E}[Q_iQ_j]= \mathbb{E}[\mathbf{W}_{Q_i}^\intercal\mathbf{Q}\mathbf{Q}^\intercal\mathbf{W}_{Q_j}]=\mathbf{W}_{Q_i}^\intercal\mathbf{C}_{QQ}\mathbf{W}_{Q_j} = 0 \end{equation}
 
  \begin{equation} \mathbb{E}[I_iI_j]= \mathbb{E}[\mathbf{W}_{I_i}^\intercal\mathbf{I}\mathbf{I}^\intercal\mathbf{W}_{I_j}]=\mathbf{W}_{I_i}^\intercal\mathbf{C}_{II}\mathbf{W}_{I_j} = 0\end{equation}
  
   \begin{equation} \mathbb{E}[Q_iI_j]= \mathbb{E}[\mathbf{W}_{Q_i}^\intercal\mathbf{Q}\mathbf{I}^\intercal\mathbf{W}_{I_j}]=\mathbf{W}_{Q_i}^\intercal\mathbf{C}_{QI}\mathbf{W}_{I_j} = 0\end{equation}
   
   Canonical correlation between $\mathbf{Q}$ and $\mathbf{I}$ can be calculated by solving the eigenvalue equations 
   \begin{equation}
   \mathbf{C}_{QQ}^{-1}\mathbf{C}_{QI}\mathbf{C}_{II}^{-1}\mathbf{C}_{IQ}\hat{\mathbf{W}}_{Q} = \rho^{2}\hat{\mathbf{W}}_{Q}
   \end{equation}
   \begin{equation}
   \mathbf{C}_{ii}^{-1}\mathbf{C}_{IQ}\mathbf{C}_{QQ}^{-1}\mathbf{C}_{QI}\hat{\mathbf{W}}_{I} = \rho^{2}\hat{\mathbf{W}}_{I}
   \end{equation}
where the eigenvalues ($\rho^{2}$) are the canonical correlations and eigenvectors ($\hat{\mathbf{W}}_{Q}$ and $\hat{\mathbf{W}}_{I}$) are the basis vectors. Only one of these equations need to be solved as the solutions are related by 
\begin{equation}
\mathbf{C}_{QI}\hat{\mathbf{W}}_{I} = \rho\lambda_{Q}\mathbf{C}_{QQ}\hat{\mathbf{W}}_{Q}
\end{equation}

\begin{equation}
\mathbf{C}_{iq}\hat{\mathbf{W}}_{Q} = \rho\lambda_{I}\mathbf{C}_{II}\hat{\mathbf{W}}_{I}
\end{equation}

where, 
\begin{equation}
\lambda_{Q}\lambda_{I} = \sqrt{\frac{\hat{\mathbf{W}}_{I}^\intercal\mathbf{C}_{II}\hat{\mathbf{W}}_{I}}{\hat{\mathbf{W}}_{Q}^\intercal\mathbf{C}_{QQ}\hat{\mathbf{W}}_{Q}}}.
\end{equation}

We can project both query and image feature vectors to the newly learned set of basis vectors to obtain new query ($\mathbf{Q}' \in \mathbb{R}^{t \times m}$) and item ($\mathbf{I}' \in \mathbb{R}^{t \times m}$) representations that are of equal dimensions. 

\begin{displaymath}
(\mathbf{Q}',\mathbf{I}') =  CCA(\mathbf{Q},\mathbf{I})
\end{displaymath}

Since both $\mathbf{Q}'$ and $\mathbf{I}'$ are equidimensional we can estimate the cosine similarity between the two. We hypothesize that since images in item features were used to learn the new space, even the query representation encapsulates the image features despite not having any images representing the query. Learning image information for queries without any images is our main contribution in this work.
 
 \begin{table*}[ht]
\centering
\begin{tabular}{|c|c|c|c|c|c|} \hline
\multicolumn{3}{|c|}{AUROC}& \multicolumn{3}{c|}{AUPRC} \\
\cline{1-6}
Baseline & Proposed Method & Gain &  Baseline & Proposed Method & Gain \\ \hline
0.5529 & 0.6182 & 11.89\% & 0.8169 & 0.8423 & 3.1\% \\ \hline
\end{tabular}
\caption{AUROC and AUPRC of the proposed method against the baseline}\label{tab:auc}
\end{table*}

\section{Experiments}
\label{sec:exp}
\subsection{Dataset}
We used total of 419,905 query-item pairs from eBay's search logs. Each of the items appeared in our search result pages for the corresponding query. Note that eBay search supports different sort types in production such as best match sort, price low to high sort among others. These query item pairs are sampled from all sort types. We had human judges label the relevance of the query item pairs. Definition of relevance is defined by eBay's internal guidelines. Judges marked these query item pairs as relevant or irrelevant. Out of the 419,905 query-item pairs, we have 328,574 pairs marked as relevant by judges and 91,331 pairs marked as irrelevant. In our dataset we have 378,888 unique items and 226,531 unique queries. All the queries occurred either in US, UK or Germany eBay site. In our dataset, queries are represented as the tokenized text and items are represented as the item title, item image and the listing category of the item. 

\subsection{Baseline}
For a baseline we used the text of the query and the item title to compute a text similarity measure. We used the listing category to derive hashed \textit{tf-idf} representation for the item title and the corresponding query as described in Section ~\ref{sec:feat}. Both query and the items feature vectors are $1000$ dimensional. We estimate the baseline cosine similarity $\mathbf{S}_{b} \in \mathbb{R}^{t}$ between the query vector $\mathbf{Q} \in \mathbb{R}^{t \times m}$ and the item title vector $\mathbf{V} \in \mathbb{R}^{t \times v}$ as :

\begin{displaymath}
\mathbf{S}_{b} = cos(\mathbf{Q},\mathbf{V})
\end{displaymath}

with \textit{t} as 419,905 and both \textit{v} and \textit{m} as 1000.

\subsection {Including Image Information Using CCA}
We will now examine how providing additional information about images makes query-item similarity a better measure for relevance. Since we formulated the query-item similarity as a cosine between query-item vectors, we will need additional information about images in both query and item vectors. However, images are only available for the item there is no image information available about the query. Therefore we will use the proposed method to include image information for both item and query in the current formulation.

We extracted $1024$ dimensional Resnet feature vectors for the item image. For doing CCA, we concatenate item image and item title features. Therefore, we have $2024$ dimensional feature vector for the item and $1000$ dimensional feature vector for the query. CCA is done over the item vector $\mathbf{I} \in \mathbb{R}^{t \times n}$ and the query vector $\mathbf{Q} \in \mathbb{R}^{t \times m}$ with \textit{t} as 419,905, \textit{n} as $2024$ and \textit{m} as $1000$. CCA learns \textit{m} basis vectors and we obtained a new item vector $\mathbf{I}' \in \mathbb{R}^{t \times m}$ and a new query vector $\mathbf{Q}' \in \mathbb{R}^{t \times m}$ by projecting the original item and query vectors onto the new learned basis vector. We then estimated proposed cosine similarity $\mathbf{S}_{p} \in \mathbb{R}^{t}$ between the new item vector $\mathbf{I}' $ and the new query vector $\mathbf{Q}'$ as shown below. 

\begin{displaymath}
(\mathbf{Q}',\mathbf{I}') =  CCA(\mathbf{Q},\mathbf{I})
\end{displaymath}

\begin{displaymath}
\mathbf{S}_{p} = cos(\mathbf{Q}',\mathbf{I}')
\end{displaymath}

\subsection{Evaluation}
\begin{figure}
\centering
\includegraphics[height=3in, width=3in]{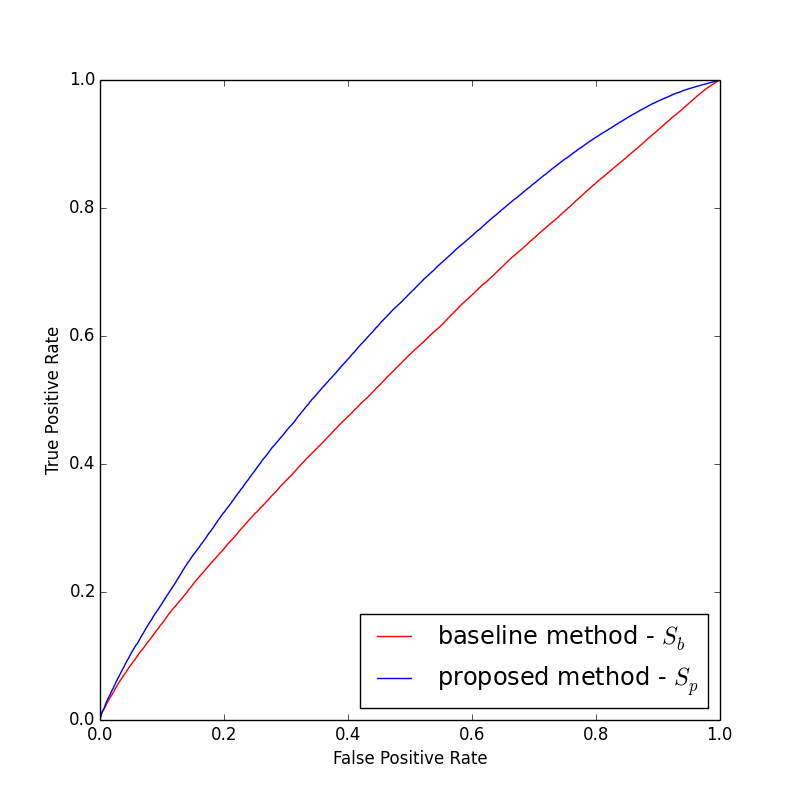}
\caption{ROC curve of the proposed method against the baseline}
\label{fig:roc}
\end{figure}

\begin{figure}
\centering
\includegraphics[height=3in, width=3in]{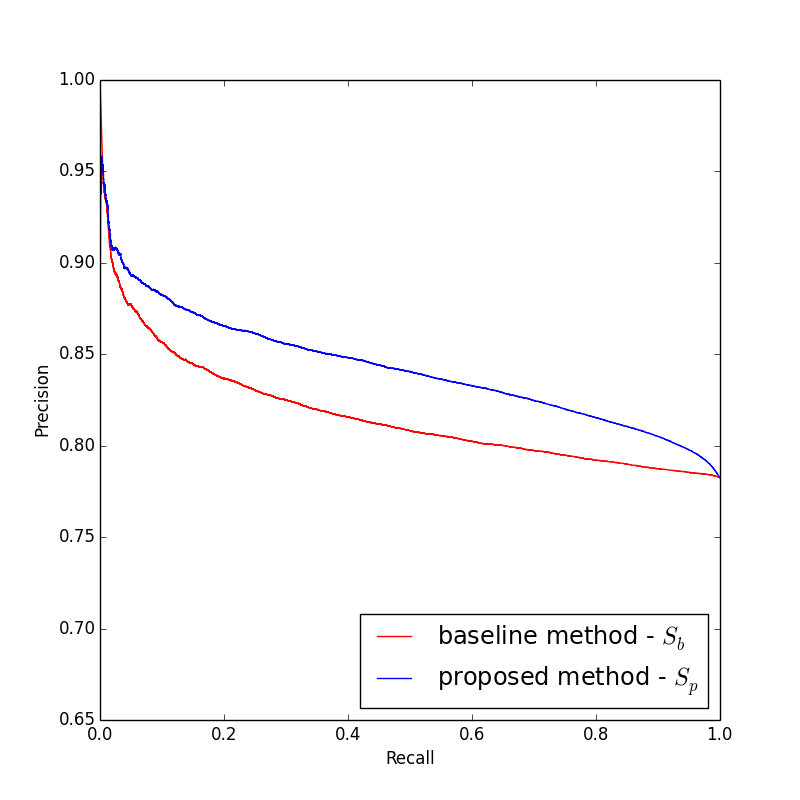}
\caption{Precision Recall curve of the proposed method against the baseline}
\label{fig:pr}
\end{figure}

We used the similarity score per query-item pair as a measure of relevance and evaluated the performance of the proposed method against the baseline method. We used Receiver Operating Characteristic (ROC) and precision-recall as our evaluation metric. Note that we have 328,574 pairs marked as relevant by judges and 91,331 pairs marked as irrelevant. Since there is some class imbalance in the dataset we decided to use both ROC and precision recall curve as our evaluation metric.  ROC and precision-recall focuses on different classes so it is useful to have both as metric. Precision-recall compares false positives to true positives while ROC compares false positives to true negatives. Therefore both metric accentuates performance of the algorithm on different classes of the data. ROC and precision recall curve of the proposed method against the baseline is shown in Figure \ref{fig:roc} and Figure \ref{fig:pr}.

The area under the ROC curve (AUROC) and the area under the precision recall curve (AUPRC) of the proposed algorithm against the baseline method is reported in Table \ref{tab:auc}. Our results demonstrate that including image information in computing the query-item similarity can significantly improve the relevance of the search engine. We propose a novel way of extracting image information for queries from item images.

\section{Conclusion}
\label{sec:conclusion}
In this paper we presented a method of deriving image information for queries using images of the items. We show 11.89\% relevance improvement in AUROC and 3.1\% relevance improvement in AUPRC on eBay search data. Our method is fully unsupervised making it highly useful for commercial information retrieval systems that have large amounts of unlabeled data at their disposal. The proposed method can be immensely helpful in obtaining image information for queries which is otherwise a non-trivial task. The proposed method has multitude of applications in ranking and retrieval as estimating similarity between query-item pairs remain a fundamental task in search.

\bibliographystyle{ACM-Reference-Format}
\bibliography{sample-bibliography}

\end{document}